\documentstyle[12pt]{article}

\begin{document}

\rightline{McGill/96-22}
\bigskip
\input epsf.tex
\centerline{\normalsize\bf SOFT INTERACTION OF HEAVY FERMIONS$^*$}

\vspace*{.6cm}
\centerline{\footnotesize C.S. LAM}
\baselineskip=13pt
\centerline{\footnotesize\it Department of Physics, McGill University}
\baselineskip=12pt
\centerline{\footnotesize\it Montreal, Quebec, Canada H3A 2T8}
\centerline{\footnotesize E-mail: Lam@physics.mcgill.ca}

\vspace*{0.9cm}
\begin{abstract}
We discuss a formula for the sum of tree diagrams of
$n$-bosons scattering from a heavy fermion. This formula can be considered
as a generalization of the eikonal formula to include non-abelian vertices.
It will be used to demonstrate the consistency of multi-meson-fermion
scattering in the large-$N_c$ limit.
\end{abstract}
\normalsize\baselineskip=15pt
\setcounter{footnote}{0}
\renewcommand{\thefootnote}{\alph{footnote}}

\def\.{\!\cdot\!}
\def\:{\!\cdots\!}
\def\[{\left[}
\def\]{\right]}
\def\({\left(}
\def\){\right)}
\def\b{$$\eqalignno}
\def\bk#1{\langle#1\rangle}
\def\bra#1{\langle#1|}
\def\bs{$\bullet$}
\def\cs{$\circ$}
\def\ds{$\diamond$}
\def\h{{1\over 2}}
\def\hsp{\hskip2cm}
\def\i#1{\item{#1}}
\def\ii#1{\itemitem{#1}}
\def\ket#1{|#1\rangle}
\def\l{\ell}
\def\o#1#2{{#1\over #2}}
\def\p{\partial}
\def\r2{\sqrt{2}}
\def\rule{\centerline{\vrule height2pt width4cm depth0pt}}
\def\srule{\centerline{\vrule height1pt width3cm depth0pt}}
\def\Tr{{\rm Tr}}
\def\ts{\triangleright}
\def\V{{\cal V}}
\def\x{\times}

\section{Introduction}

This work is carried out in collaboration with Keh-Fei Liu of the University
of Kentucky$^1$. The central piece is a tree-level
formula for the sum of the $n!$ diagrams describing the scattering of $n$
bosons from a heavy fermion.
This formula  
is kinematical and combinatorial,
valid for any coupling and internal quantum numbers.  It relies on the presence
of a heavy fermion,
just like the heavy-quark effective theory.
It will be used
to demonstrate the consistency of meson-baryon scattering in
large-$N_c$ QCD$^2$ for
an {\it arbitrary} number of mesons, under
certain criteria.
We are in the process of extending it
to study the consistency of large-$N_c$ QCD in loops,
baryon-baryon scattering,  and heavy quarks.

This formula bears some superficial similarity
to formulas for soft-pion productions$^3$, in that multiple
commutators appear in both. However, the present formula does not
rely on chiral symmetry, nor current algebra. As stressed before,
it is kinematical and combinatorial, it is generally valid under
no particular dynamical assumptions.
Its validity depends on the smallness of all energy scales
compared to the heavy-fermion mass $M$, and not some dynamical
quantity like $f_\pi$. If we need an analogy with known formulas, then
it is much more appropriate to think of it as a non-abelian
generalization of the eikonal formula$^4$.

\section{Sum of Feynman Diagrams}

Fig.~1 depicts one of the $n!$ tree diagrams for the process
$$F'(p')\to F(p)+B_1(k_1)+B_2(k_2)+\cdots B_n(k_n)\ ,\eqno(1)$$
in which an initial fermion $F'$ emits $n$ bosons $B_i$ to become a
final fermion $F$. The momenta of
the particles are enclosed in parentheses. If a boson in (1)
is moved from right to left, we will get an inelastic boson-fermion
scattering
process. Both
possibilities are considered using the notation of (1).
The initial and the final fermions are assumed to have a heavy mass $M$:
$p^2=(p')^2=M^2$.

\begin{figure}
\vskip -0 cm
\centerline{\epsfxsize 4.7 truein \epsfbox {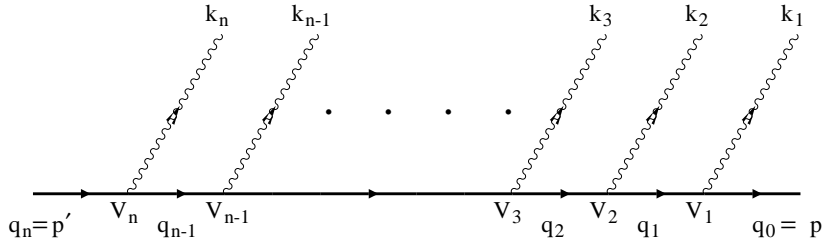}}
\nobreak
\vskip -8.5 cm\nobreak
\vskip -1 cm
\caption{A tree diagram. Solid and Wavey lines are respectively fermions
and bosons.}
\end{figure}

We shall use the notation $T_a=[t_1t_2t_3\cdots t_n]$
to denote a tree graph obtained from Fig.~1 by permuting the $n$
boson lines. The index $a$ runs from 1 to $n!$, and each $t_i$ takes
on a different value between 1 and $n$.
The numbers between the
square brackets indicate the order of the boson lines from final to initial
states.
In this language Fig.~1 is denoted by $T_1=[123\cdots n]$.

One encounters these diagrams in calculating the emission of soft
photons from a (comparatively)
heavy fermion. The result, summed over the $n!$ diagrams, is described by the
familiar `eikonal formula'$^4$, according to which the $n$ photons may be
treated as
being emitted independently from the heavy source. Note that such factorization
does not occur in each Feynman diagram so the sum is vastly
simpler than the individual diagrams.
Since we need to obtain a non-abelian generalization of the eikonal
formula, it is useful first to recall how the abelian case is obtained.

The $i$th propagator of Fig.~1 is given by
$$S_i
={M_i+\gamma q_i\over M_i^2-q_i^2-i\epsilon}\ ,
\quad q_i=p+\sum_{j=1}^ik_j\ .   \eqno(2)$$
For other permuted diagrams $k_j$ should be replaced by $k_{t_j}$, but
in every case we have $q_0=p$ and $q_n=p'$.
Let $\omega=p\.k_i/M$ be the energy of the $i$th boson in the rest frame
of the final fermion. Our central assumption is  $\omega_i\ll M$,
{\it i.e.,} that the interaction
is soft compared to the mass of the fermion.
Then the numerator of  propagator (2) may be approximated by $M+\gamma p$
provided the intermediate masses $M_i$ are not  too different from $M$:
$\Delta M_i\equiv (M_i^2-M^2)/2M\ll M$. In that case (2)
can be approximated by 
$$S_i={\cal P}D_i\ ,\eqno(3)$$ where
$$D_i={1 \over \Delta M_i-W_i-i\epsilon}\ ,\eqno(4)$$
$W_i=\sum_{j=1}^i\omega_{j}$ is the total boson energy preceeding
that propagator, and ${\cal P}=\h(1+\gamma p/M)$ is a projection operator.
With this approximation, the vertex factor $\gamma^\mu$ can
be replaced by the velocity $v^\mu=p^\mu/M$ of the fermion, because
$${\cal P}\gamma^\mu{\cal P}=v^\mu{\cal P}\ .\eqno(5)$$

For the abelian eikonal formula,
a single fermion without resonance is considered,
so $\Delta M_i=0$.
The {\it off-shell} (for line $p'$) photon-emission amplitude from
 Fig.~1 is then
$$A_0^*[T_1]=\prod_{i=1}^n\(2ev\.\epsilon(k_i)D_{i0}\)\ ,\eqno(6)$$
where the subscript 0 is there to remind us of the condition $\Delta M_i=0$,
and the asterisk denotes `off-shell'. Note that $W_i$ in $D_i$ contains
a sum of meson energies so there is no factorization here.

With a certain amount of combinatorial algebra, it can be shown$^4$ that the
total off-shell emission amplitude is
$$A_0^*=\sum_{a=1}^{n!}A^*[T_a]=\prod_{i=1}^n\(-{ev\.\epsilon(k_i)
\over p\.k_i}\)\ .\eqno(7)$$
This eventual factorization of the boson variables
then allows the conclusion
that the $n$ photons are emitted independently.

For {\it on-shell} amplitudes, the condition $M^2-(p')^2=0$ is equivalent to
the energy-conservation condition
$${1\over M}p\.\sum_{i=1}^nk_i=\sum_{i=1}^n\omega_i=0\eqno(8)$$
in the approximation $\omega_i\ll M$.
The on-shell amplitude $A_0$ is equal to the off-shell amplitude $A^*_0$
with the
$n$th propagator removed, so it is
$$A_0\sim (M^2-p^{'2})A^*_0=0\eqno(9)$$
A more careful
calcluation shows that $A_0$ is actually
proportional to the products of $\delta(\omega_i)$$\ ^4$. Formulas
(7) and (9) will be referred to as the {\it abelian eikonal formula}.

We shall now proceed to consider the non-abelian generalization of this
formula. In that case, the on-shell amplitude from Fig.~1 is
$$
A[T_1]=\bra{p\lambda}\V_1S_1\V_2S_2\V_3\cdots \V_{n-1}S_{n-1}
\V_n\ket{p'\lambda'}\eqno(10)$$
where $\V_i$ is the $i$th vertex attached to the boson $B_i$,
$S_i$ is the propagator in (2), and $\lambda, \lambda'$ represent
helicities and other internal quantum numbers. We allow in this non-abelian
version multi-channel configurations where both $\V_i$ and $\Delta M_i$ should
be regarded as matrices connecting fermions of various helicities and
internal quantum numbers. These matrices do not commute and it is in this
sense that the amplitude is {\it non-abelian}. Within the same heavy-fermion
approximation, the numerator $M+\gamma q_i$ in (2) may be replaced by its
on-shell value and written as $\sum_{\lambda_i} u_{\lambda_i}(q_i)\bar
u_{\lambda_i}(q_i)$. In this way (10) can be written as
$$
A[T_1]= V_1D_1V_2D_2V_3\cdots V_{n-1}D_{n-1}V_n
\eqno(11)$$
provided we take
$$V_i=\bk{q_i\lambda_i|\V_i|q_{i-1}\lambda_{i-1}}\eqno(12)$$
and $D_i$ defined in (4).
For the permuted diagrams, $V_i$ and $\Delta M_i$ are to be replaced by
$V_{t_i}$ and $\Delta M_{t_i}$.

Eq.~(11) will be our starting point for further analyses. As discussed above,
it is valid to the leading order of $\omega_i/M$. However, if we
interpret $A[T_1]$ as the pole-part of the amplitude, then (11) is {\it exact}
provided the bosons are massless
and the boson momenta are all parallel to one another
 ({\it `forward scattering'}). This is
so because under that condition $k_i\.k_j=0$ so the expression $D_i$ in
(3) and (4) is
exact. So is eq.~(8).
Moreover, by taking the pole part of the amplitude,
the numerator of (2)  should be taken on shell (the residue of the poles)
so the vertices can be rigorously replaced by their helicity matrix elements as
in (12).

Given that, it is more satisfying to regard
(11) in this way, as an {\it exact} expression for the pole part of
the forward massless scattering amplitude, so that an
{\it exact} non-abelian formula for the sum can be obtained.
At the end, we might want to put in boson masses, consider
non-forward scattering, and the non-pole part of the tree amplitude.
These lead to corrections
of order $\omega_i/M$, so the formula would still be valid to
the zeroth order of this small parameter.

To proceed further, let us first continue to ignore the possibility of
resonances and put
all $\Delta M_i=0$. Eq.~(11) then becomes
$$A_0[T_1]=V[T_1]D_0[T_1]\ ,\eqno(13)$$
where
$$V[T_1]=V_1V_2V_3\cdots V_{n-1}V_n\ ,\eqno(14)$$
and
$$D_0[T_1]=\prod_{i=1}^{n-1}D_{i0}=\prod_{i=1}^{n-1}{1\over
-W_i-i\epsilon}\ .\eqno(15)$$
The sum over all diagrams
$A_0=\sum_{a=1}^{n!}A_0[T_a]$
can then be obtained by a combinatorial argument$^1$ to be
$$A_0=\sum_{b=1}^{(n-1)!}V^{MC}[T'_bn]D_0[T'_bn]\ , \eqno(16)$$
with the following notations. $[T'_bn]=[t'_{b_1}t'_{b_2}\cdots
t'_{b_{n-1}}n]$ is
the
subset of trees $T_a$ with the boson $B_n$ fixed at the end, and the sum in
(16) is over the $(n-1)!$ permutations of the remaining boson lines.
$V^{MC}$ stands for `multiple commutator'
$$V^{MC}[T'_bn]=[V_{b_1}[V_{b_2}[V_{b_3}\ \:\ [V_{b_{n-1}},V_n]\ \:\ ]]]\ .
\eqno(17)$$
Line $n$ is singled out in (16) but we could have singled out any other line
instead.
Note that the derivation of (16) is not trivial, as can be seen from the
fact that each multiple commutator contains $2^{n-1}$ terms so (16) contains
a total of $(n-1)!2^{n-1}$ terms, whereas (15) where this comes from contains a
total of only $n!$ terms.

\begin{figure}
\vskip -0 cm
\centerline{\epsfxsize 4.7 truein \epsfbox {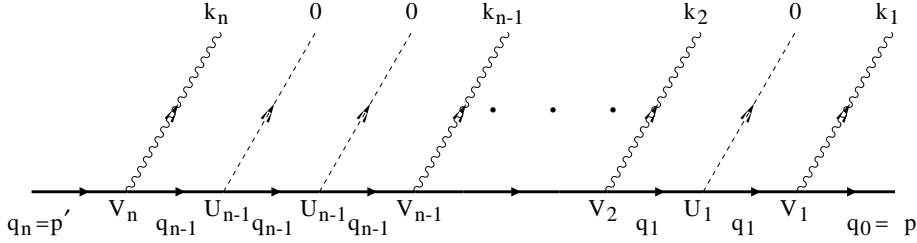}}
\nobreak
\vskip -8.5 cm\nobreak
\vskip -1 cm
\caption{Same as Fig.~1 but with spurious bosons (dashed lines)
put in to take care of resonance-mass corrections.}
\end{figure}

Finally let us see how the formula changes when resonances are included
to make $\Delta M_i\not=0$. In that case, instead of
$D_{i0}$, we should use the full $D_i$ in (4). By using the expansion
$$D_i=D_{i0}\sum_{\ell_i=0}^\infty \(-\Delta M_i D_{i0}\)^{\ell_i}\ ,
\eqno(18)$$
we can 
cast this amplitude into sums of amplitudes without resonances
by introducing spurious bosons.
Specifically, for a given
$\ell=\sum_{i=1}^{n-1}\ell_i$, the expanded amplitude can be represented
by Fig.~2, which differs from Fig.~1 by having $\ell$ spurious bosons
(indicated by dotted lines) added. These dotted lines
must appear between the first and the last
wavey lines, but otherwise they are to be inserted in all possible
ways. The dotted lines
 carry zero energy and a vertex factor $U_{t_i}\equiv -\Delta M_{t_i}$, where
$t_i$ is the original (wavey line)
boson immediately to the right of the dotted line
concerned. All dotted lines between two consecutive wavey lines carry
the same zero energy and possess the same vertex, so they are to be thought of
as identical particles. In the presence of resonances,
the sum over all diagrams are again given by (16), but now modified to
describe Fig.~2 instead of Fig.~1, and a sum over all $\ell$
must also be taken. The permutation in (16)
is similarly modified to include all permutation of non-identical bosons in
Fig.~2. Inspite of the many sums present, the important thing is that
it remains to be given by multiple commutators of the $V_i$'s and the
$U_i$'s.

\section{Large-$N_c$ QCD}

We can use (16) to demonstrate the consistency of
inelastic meson-baryon scattering amplitude in large-$N_c$ QCD.
The requirment of consistency for $n=2$ and 3 led to very interesting
physical predictions$^2$. For an arbitrary $n$, to our
knowledge the consistency has not been demonstrated and we shall use (16)
to show it.

Let me review briefly the problem encountered in the limit of large-$N_c$.$^2$
\  The baryon mass $M$ in that limit is of order $N_c$, so from (2)--(4) the
baryon propagator is of order 1. However, the meson-baryon Yukawa coupling
constant grows like $\sqrt{N_c}$, so a diagram like Fig.~1 will be of order
$N_c^{n/2}$. On the other hand, the complete
$n$-meson baryon amplitude is
known to go down with $n$ like $N_c^{1-n/2}$. There is then a discrepency of
$n-1$ powers of $N_c$ between an individual diagram and the complete amplitude.
How can one effect such a huge cancellation between the individual diagrams
in the sum?
This is the consistency problem mentioned above. The problem of course gets
worse as $n$ increases.

For $n=2$ and 3, a multiple-commutator
formula similar to (16) had been used to demonstrate
the consistency$^2$. It proceeds by showing that though individual
vertices $V_i$ are of order $\sqrt{N_c}$ so the product of $n$
vertices behave like $N_c^{n/2}$, the multiple commutators of these vertices
in fact behaves like $N_c^{1-n/2}$, as desired. The same proof works for
an arbitrary $n$ as long as a formula like (16) is available to convert
the sum of diagrams into multiple commutators.

To understand why multiple commutators lead to such an enormous cancellation,
we have to go back to study how a large-$N_c$ baryon is made
of and how it couples. A baryon in this framework is a color-singlet object
composed of $N_c$
quarks in the S state. Different baryons arise from different spin and flavor
excitations, but the wave function must be symmetric in the spin and flavor
variables of the $N_c$ quarks. Spin is described by the
rotational group $SU(2)_J$ so the spin of a baryon is
represented by the Young tableau in Fig.~3(a). Each box in
the tableau represents
a quark.
Those columns with two boxes are spin singlets,
and those columns with  one box are spin doublets symmetrical under
permutation of these columns,
so the spin of the baryon is $J$ if there are $2J$ single-box columns.
Fig.~3(b) is for the flavor group. The tableau is identical to the one in 3(a)
to enforce the symmetry of the wave function. This
time each box represents the flavor state
of a quark. Under isotopic-spin symmetry the
flavor group is $SU(2)_F$. The total isospin of the baryon is $I=J$ because
there are $2J=2I$ single-box columns in 3(b). Putting these together we
conclude that the allowed baryon resonances have $I=J=\h, 1, {3\over 2}, \:,
{N_c\over 2}$.

\begin{figure}
\vskip -4 cm
\hskip3cm\centerline{\epsfxsize 4.7 truein \epsfbox {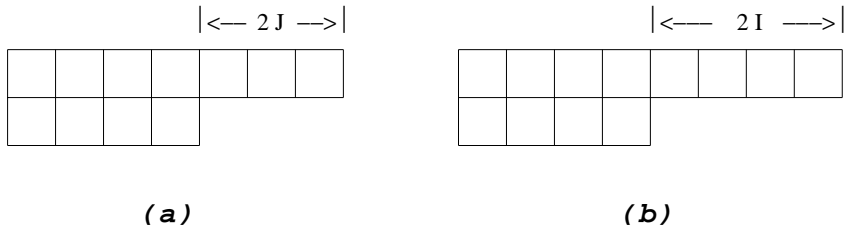}}
\nobreak
\vskip -6 cm\nobreak
\vskip -2.5 cm
\caption{Young tableaux for (a) $SU(2)_J$ and (b) $SU(2)_F$.}
\end{figure}

On the other hand, if we assume a flavor symmetry of $SU(3)_F$,
then the two-box columns in 3(b) give rise to a ${\bf \overline 3}$
representation while the single-box columns give rise to a ${\bf 3}$
representation. If there are $p$ single-box columns and $q$ two-box columns
in 3(b), so that $N_c=p+2q$, then the $SU(3)_F$ representation of the baryon
is $(p,q)$ (where $(1,1)$ is the octet, (3,0) is the decuplet, (2,2)
is the 27-plet, etc). For large $N_c$, one or both of $(p,q)$ must be large,
and we are then talking about $SU(3)_F$ representaions too large to appear in
Nature.
Consequently it is useless to talk about large $N_c$ as an approximation to
Nature  if we insist on $SU(3)_F$ symmetry as well as the identical-particle
symmetry of the baryonic
wave function. In that sense large
$N_c$ hates $SU(3)_F$, which makes the sucessful
broken-$SU(3)_F$ predictions from the large-$N_c$
analyses$^2$ even more intriguing.
To consider large $N_c$ as a useful tool we must
go back to $SU(2)_F$ symmetry. We may still include strange and other
quarks, but in no way they should be regarded as identical to the up
and down quarks. New quarks of each species
 in the baryon are to be represented by a
new tableau in Fig. 3(a), with $2K$
single-box columns if that is the number of new quarks of that species.
They carry a total spin of $K$.
There will be
no corresponding tableau for these new quarks in 3(b)
because they are strong-isospin
singlets. Consequently the total spin carried by the up and down
quarks is still $I$, so the
allowed spin $J$ of the baryon is obtained by adding $I$ and $K$ vectorially,
{\it i.e.,} it ranges from $|I+K|$ to $|I-K|$.

We are now ready to describe how the baryon matrix elements for the vertices
are computed. Let $q^\dagger,q$ be free-quark creation and annihilation
operators. Let us abbrviate a one-body (free) quark operator as
$$q^\dagger\Gamma q\equiv \{\Gamma\}\ ,\eqno(17)$$
where $\Gamma$ is either the (spin) Pauli matrices $\sigma^i$,
the flavor matrices $t^a$,  their products, or ${\bf 1}$. For two flavors,
$t^a=\tau^a$
are the Pauli matrices, and for three flavors, they are the Gell-Mann
matrices $t^a=\lambda^a$. The use of $SU(3)_F$ quark operators and matrices
does not imply that the baryon wavefunction is $SU(3)_F$ symmetric.

Any operator ${\cal O}$ can always be written as sums of products of
operators of the form (17), the only constraint being that
they must have the same quantum numbers as ${\cal O}$. A product of $m$
one-body operators will be referred to as an $m$-body operator,
and it can be shown$^2$ that an explicit factor $F_m=N_c^{1-m}$ must accompany
the presence of an $m$-body operator.
We need to find out the $N_c$-dependence of the baryon matrix elements of
$\{\Gamma\}$ and its products. We are interested only in baryons of low
$J$ and  $I$ which exist in Nature, hence
most of the columns (of order $N_c$) in Fig.~3 are two-box columns.
They constitute singlet spins and singlet isospins and contribute zero to
 matrix
elements of $\Gamma=\sigma^i$ and $\tau^a$. For these $\Gamma$, the baryon
matrix element
$$\bk{\Gamma}=\bk{q_i\lambda_i|\{\Gamma\}|q_{i-1}\lambda_{i-1}}\eqno(18)$$
comes only from the single-box columns so it
is of order 1. For $\Gamma={\bf 1}$, or $\sigma^i\tau^a$, or if it involves
matrices connecting the up/down quarks to other quarks,
the two-box columns
do contribute so the matrix element $\bk{\Gamma}$ is generally of order $N_c$,
the order of the number of columns.
Similar arguments show that the matrix element $\bk{\{\Gamma\}^m}$
of an $m$-body operator is at most of order $N_c^m$, so with the
factor $F_m$ in front, the matrix element of every term of ${\cal O}$
is at most of order $N_c$ (assuming that there are no explicit factors
of $N_c$ accompanying the definition of ${\cal O}$).

Though $\bk{\{\Gamma\}^m}$ is generally of order $N_c$,
it might be as suppressed as
$N_c^{1-m}$
 if all the $\Gamma$'s are made up of $\sigma^i$'s and
$\tau^a$'s. When this occurs important phenomenological consequences
often follow$^2$. For example, if all the $m>1$ -body operators are
suppressed, then the dominant contribution comes from the one-body operator,
but that is just treating the quarks inside the baryon as if they were free,
so this gives rise to a large-$N_c$ justification of the quark model.
Another example has to do with the mass of a baryon. It can be computed from
the matrix elements of the operators $\{1\}, \{\vec J\}\.
\{\vec J\}/N_c$, etc., so it is of the form $M_i=aN_c+bJ(J+1)/N_c$. The mass
split $\Delta M_i$ is not only small compared to $M$ as demanded by
the approximations needed to reach (16), it is actually of order $N_c^{-1}$
and not just order 1. Like the Skyrme model, or the strong-coupling model,
the baryonic resonances are rotational levels very closely
packed.

It is now easy to see why (16) leads to the enormous
cancellation between individual Feynman diagrams needed for the consistency.
From free-quark commutation relation, the
commutator of two 1-body operators is again a 1-body operator, and the
commutator of an $m$-body operator with an $m'$-body operator is an
$(m+m'-1)$-body operator. As far as matrix elements are concerned, this
means that each time we commute we lose one power of $N_c$, so
all together $(n-1)$ powers of $N_c$ are lost by the multiple commutator
in (16), and this is precisely the amount needed to match the discrepency
between individual and the sum of meson-baryon scattering diagrams.

If baryon resonances are included, (16) can still be used with modifications,
as explained. The argument above valid for $\Delta M_i=0$ will still be valid
if the additional vertices $U_i=-\Delta M_i$ are of order 1. As we saw above,
this is actually of order $N_c^{-1}$ so there is no problem when baryon
resonances are present either.

In this way we prove the consistency of the large-$N_c$ amplitudes
{\it if} (16) is used, with or without baryonic resonances. Eq.~(16) is exact
for the pole part of
the massless forward scattering amplitude. When boson masses are taken
into account and the forward scattering restriction is relaxed,
eq.~(16) is only approximate. As such cancellations arising from the
multiple commutators are still present, but (16) alone is not sufficient
to demonstrate the {\it complete} cancellation of the $(n-1)$ powers of $N_c$.
We shall not discuss this more difficult problem here.

\section{Acknowledgements}
I am grateful to Y.J. Feng for drawing the diagrams.

\end{document}